\documentclass[conference]{IEEEtran}
\IEEEoverridecommandlockouts
\usepackage{cite}
\usepackage{amsmath,amssymb,amsfonts}
\usepackage{algorithmic}
\usepackage{graphicx}
\usepackage{textcomp}
\usepackage{xcolor}

\usepackage{hyperref}
\usepackage{titlesec}
\titlespacing*{\subsection}{0pt}{4pt}{3pt} 

\def\BibTeX{{\rm B\kern-.05em{\sc i\kern-.025em b}\kern-.08em
    T\kern-.1667em\lower.7ex\hbox{E}\kern-.125emX}}

\begin{document}

\title{Building Trust in the Skies: A Knowledge-Grounded LLM-based Framework for Aviation Safety\\
}

\author{
    \IEEEauthorblockN{Anirudh Iyengar*, Alisa Tiselska*, Dumindu Samaraweera, \textit{Senior Member, IEEE}, and Hong Liu}
    \IEEEauthorblockA{Department of Mathematics, Embry-Riddle Aeronautical University, Daytona Beach, FL  \\ \{iyengara, tiselska\}@my.erau.edu, \{samarawg, liuho\}@erau.edu}

    \thanks{\IEEEauthorrefmark{1}These authors contributed equally to this work.} 
}

\maketitle

\titlespacing*{\section}{0pt}{0.5\baselineskip}{0.3\baselineskip}
\titlespacing*{\subsection}{0pt}{0.3\baselineskip}{0.1\baselineskip}

\begin{abstract}

    The integration of Large Language Models (LLMs) into aviation safety decision-making represents a significant technological advancement, yet their standalone application poses critical risks due to inherent limitations such as factual inaccuracies, hallucination, and lack of verifiability. These challenges undermine the reliability required for safety-critical environments where errors can have catastrophic consequences. To address these challenges, this paper proposes a novel, end-to-end framework that synergistically combines LLMs and Knowledge Graphs (KGs) to enhance the trustworthiness of safety analytics. The framework introduces a dual-phase pipeline: it first employs LLMs to automate the construction and dynamic updating of an Aviation Safety Knowledge Graph (ASKG) from multimodal sources. It then leverages this curated KG within a Retrieval-Augmented Generation (RAG) architecture to ground, validate, and explain LLM-generated responses. The implemented system demonstrates improved accuracy and traceability over LLM-only approaches, effectively supporting complex querying and mitigating hallucination. Results confirm the framework’s capability to deliver context-aware, verifiable safety insights, addressing the stringent reliability requirements of the aviation industry. Future work will focus on enhancing relationship extraction and integrating hybrid retrieval mechanisms.
\end{abstract}

\begin{IEEEkeywords}
    Large Language Models, Knowledge Graphs, Retrieval-Augmented Generation, Trustworthy AI, Semantic Embedding, Langraph
\end{IEEEkeywords}

\section{Introduction} \label{introduction}
Aviation is an ultra-safe, safety-critical domain in which even minor errors can lead to catastrophic outcomes. Large Language Models (LLMs) offer significant potential to augment human expertise by automating labor-intensive tasks such as analyzing large volumes of incident reports, interpreting complex regulatory documents (e.g., FAA Advisory Circulars, EASA AMCs), and providing decision support to safety managers, controllers, and maintenance personnel. By transforming unstructured text into actionable insights, LLMs can enhance situational awareness, accelerate incident analysis, and support predictive risk assessment.

However, the direct deployment of LLMs in safety-critical aviation workflows is inherently risky. LLMs are known to produce factual inaccuracies, hallucinations, and unverifiable outputs, with no explicit linkage to authoritative sources. In regulated environments, where traceability, auditability, and compliance are mandatory, such “black-box” behavior is unacceptable. An ungrounded recommendation, particularly one that misinterprets regulations or invents procedures, can propagate systemic risk, rendering standalone LLMs unsuitable as primary decision agents in aviation safety. Recognizing these opportunities and risks, the Federal Aviation Administration (FAA) has outlined a roadmap for AI safety assurance to guide responsible development and deployment within the aerospace system \cite{FAA_AI_Roadmap_2024}, \cite{Pham_REDAC_NAS_2024}.

Knowledge Graphs (KGs) provide structured, explicit, and auditable representations of domain knowledge, enabling deterministic reasoning and consistency checking. In aviation safety, KGs can encode relationships among aircraft types, components, failure modes, regulatory standards, incidents, and operational procedures. This structure ensures traceability and verifiability, key requirements for regulatory oversight and accident investigation. However, KGs have traditionally suffered from high manual construction costs and limited adaptability, making them difficult to maintain in the face of continuously evolving operational data and regulations.

Thus, these complementary strengths and weaknesses motivate a hybrid approach: combining the linguistic flexibility and extraction capabilities of LLMs with the structured, verifiable reasoning of KGs. This work addresses the gap between the transformative potential of LLMs and the stringent safety assurance requirements of aviation systems. We propose a tightly coupled, end-to-end framework that integrates LLMs and Knowledge Graphs to enable trustworthy, explainable, and regulation-aware AI assistance. Our core contribution is a dual-phase pipeline tailored for high-reliability environments. In the first phase, LLMs are used to automatically construct and continuously update an Aviation Safety Knowledge Graph (ASKG) from heterogeneous sources, including incident reports, regulatory documents, and maintenance data. In the second phase, the curated KG is used to ground, validate, and explain LLM outputs within a Retrieval-Augmented Generation (RAG) architecture. This creates a closed-loop system in which LLM-generated insights are constrained by authoritative, structured knowledge, ensuring traceability and verifiability.

By integrating graph-based reasoning with LLM inference, the proposed framework delivers context-aware, explainable safety recommendations while supporting real-time validation and regulatory compliance. This approach directly addresses critical limitations of existing AI-assisted safety systems and advances the safe adoption of AI in aviation.

The remainder of this paper is structured as follows. Section II reviews background and related work on LLMs in safety-critical domains, KGs in aviation, and hybrid AI systems. Section III details the methodology of the proposed LLM-KG integrated pipeline. Section IV describes the system implementation. Section V presents and discusses the experimental results. Section VI examines the limitations and outlines future research directions. Finally, Section VII concludes the paper.

\section{Background and Related Work}
This section positions our work within the broader landscape of AI applications in safety-critical domains, specifically examining the individual trajectories of LLMs and Knowledge Graphs in aviation, before reviewing emergent hybrid approaches and identifying the research gap our pipeline addresses.

\subsection{Large Language Models in Safety-Critical Domains}
The application of LLMs in high-reliability industries has progressed from general-purpose chatbots to specialized assistants. In aviation, early explorations focused on automated text analysis of mandatory occurrence reports (e.g., ASRS, ECCAIRS) to categorize events and identify latent risk patterns, a task traditionally requiring extensive manual review by safety experts \cite{zhang2022automated}. More advanced applications involve using LLMs for regulatory querying, where models like GPT-4 are prompted to interpret complex documents like ICAO Annexes or Part-121 operating rules \cite{chen2023regulatory}. However, rigorous evaluations reveal persistent, critical shortcomings. Studies demonstrate that even fine-tuned domain-specific LLMs exhibit hallucination rates between 15-25\% when answering technical queries, generating plausible but incorrect references to non-existent regulations or aircraft system limitations \cite{miller2024hallucination}. This is compounded by a temporal knowledge cutoff, leaving models unaware of recent Airworthiness Directives (ADs) or incident bulletins. Furthermore, LLMs lack provenance attribution, making it impossible for a safety engineer to validate the source of a recommendation, a fundamental violation of aviation's \emph{just culture} and traceability requirements mandated by standards like DO-178C and ISO/IEC/IEEE 8800. Consequently, while promising for initial triage or summarization, LLMs are not considered trustworthy for closed-loop decision-making without a verification mechanism.

\subsection{Knowledge Graphs in Aviation and Safety Management}
Knowledge Graphs offer a paradigm shift from unstructured data to semantically connected, machine-readable knowledge. In aviation safety, foundational work has established domain-specific ontologies such as the Aviation Safety Ontology (ASO), which formalizes concepts like Hazard, RiskControl, OrganizationalFactor, and their interrelations \cite{stroe2021aviation}. These structured models enable powerful applications: KGs have been used for causal chain analysis, where incident data is mapped into a graph to visually trace a primary failure (e.g., sensor fault) through intermediate events to an ultimate outcome (e.g., controlled flight into terrain) \cite{wang2023causal}. They also support regulatory compliance checking by linking operational procedures to specific regulatory clauses, allowing for automated gap analysis \cite{kim2022regulatory}.

However, the predominant challenge has been knowledge acquisition. Building and maintaining these KGs has historically been a manual or semi-automated process reliant on rule-based extractors and significant domain expert labor, making them expensive to scale and slow to update. They often become static \textit{snapshots} that fail to incorporate the continuous stream of new incident reports, `service bulletins', and `lessons learned' from Safety Management Systems (SMS). This weakness limits their utility in dynamic, real-time decision support scenarios.

\begin{figure}[htbp]
    \centering
    \includegraphics[width=0.90\linewidth]{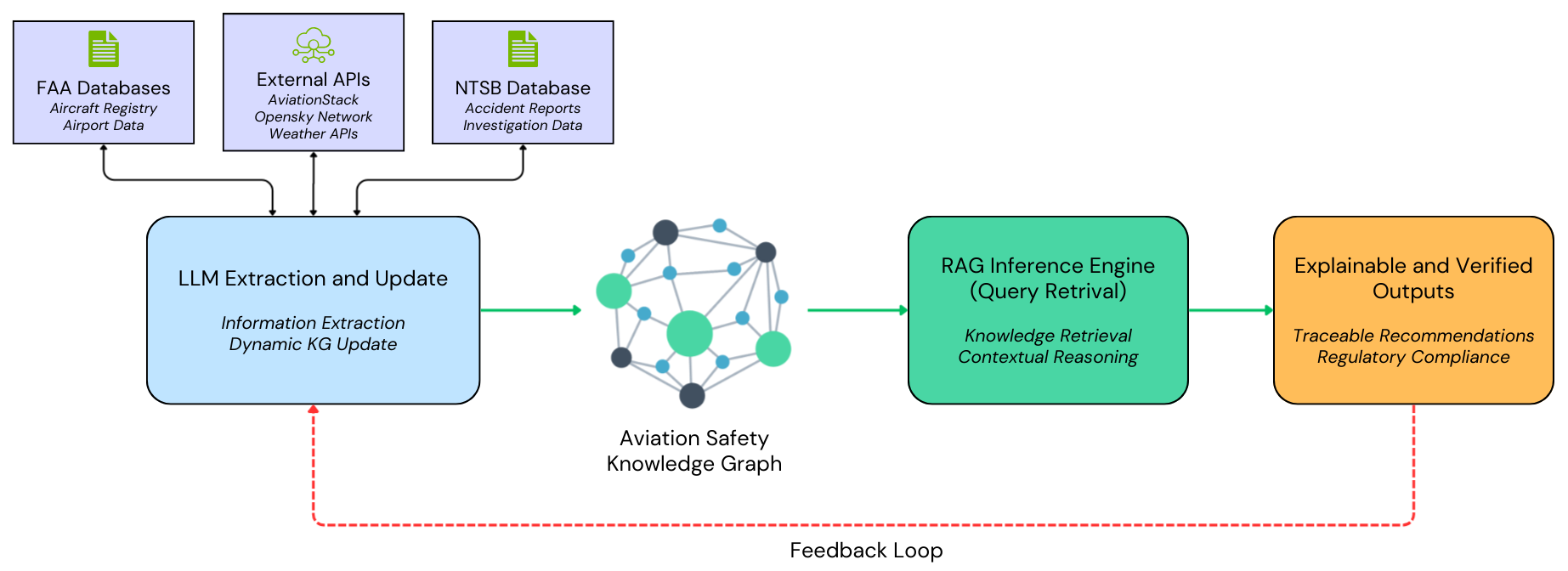}
    \caption{High-level system architecture of the proposed framework, unifying LLM-driven KG construction and KG-grounded reasoning into a single, coherent architecture.}
    \label{fig:Proposed_design}
\end{figure}

\subsection{Hybrid Approach of Combining LLMs and KGs}
The complementary strengths and weaknesses of LLMs and KGs have urged interest in hybrid architectures. The most prominent pattern is Retrieval-Augmented Generation (RAG), where a retriever fetches relevant text chunks from a corpus to ground an LLM's generation. In generic domains, this reduces hallucination by providing context. However, in technical safety domains, standard RAG has limitations: it retrieves text passages but not structured facts, and the retrieved context itself may be unverified, propagating errors from source documents \cite{lewis2020retrieval}.

A more structured evolution is Graph RAG, where retrievers query a knowledge graph to fetch connected subgraphs of entities and relations, providing the LLM with rich, semantically organized context \cite{edge2024graph}. Concurrently, research into LLM-powered KG construction has emerged. LLMs are now used as zero-shot or few-shot information extractors to populate graph schemas from text, significantly reducing manual curation effort \cite{pan2023zeroshot}. Other works explore using LLMs for graph reasoning, translating natural language questions into graph query languages like Cypher or SPARQL \cite{sun2023natural}.

\subsection{Identified Research Gap}
\begin{figure}[htbp]
    \centering
    \includegraphics[width=0.99\linewidth]{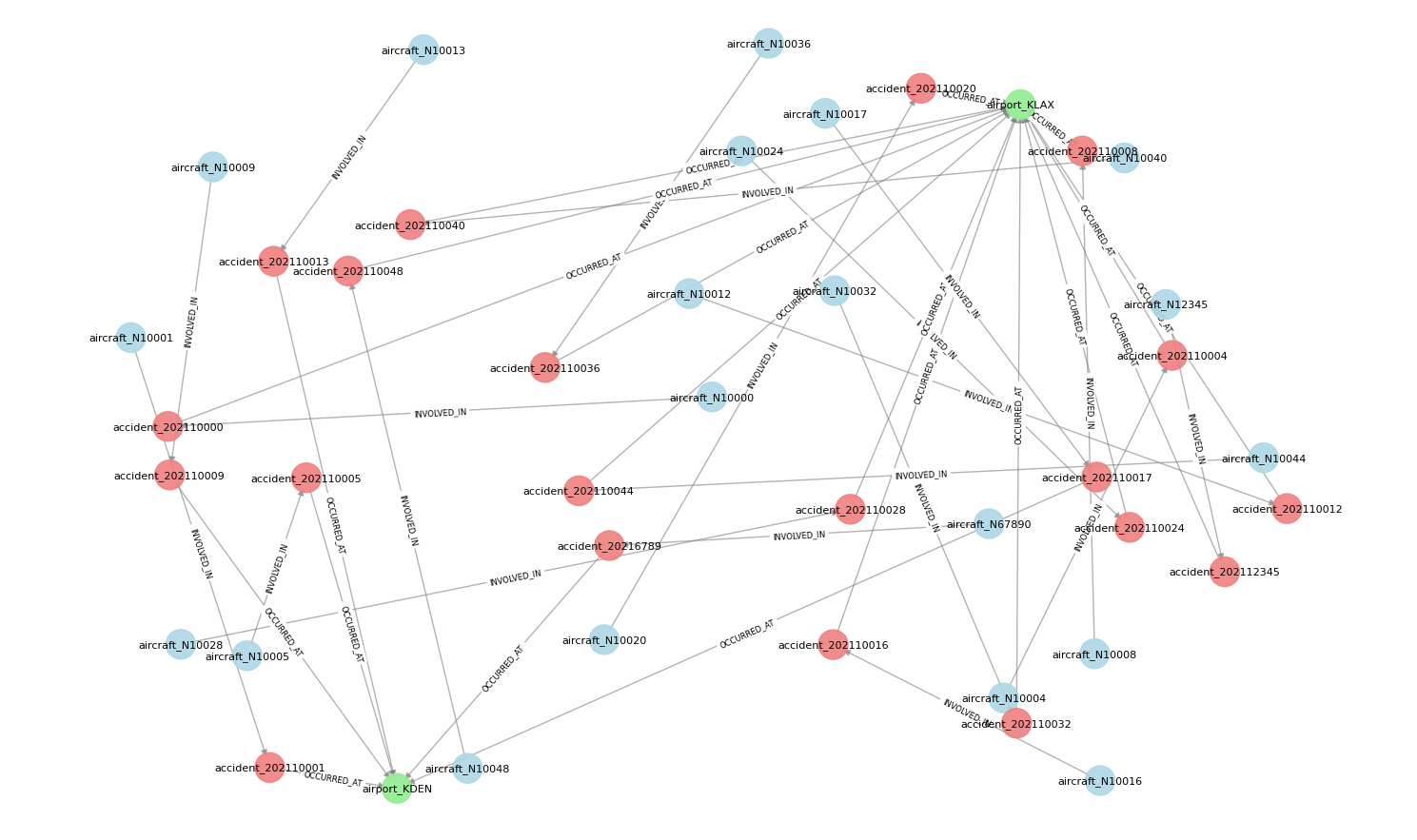}
    \caption{A sub-knowledge graph generated by the pipeline from a sample set of preliminary data.}
    \label{fig:generated_KG}
\end{figure}

Despite these advances, a critical gap remains. Existing hybrid approaches are largely modular and one-directional. They either (1) use a static, pre-built KG to augment an LLM (Graph RAG), or (2) use an LLM to build a KG offline. There is a lack of an integrated, closed-loop, end-to-end pipeline where the KG is dynamically constructed and maintained by the LLM from operational data streams, and in turn, that same continuously evolving KG is used to rigorously ground, validate, and explain the LLM's real-time outputs. Such a system would create a self-reinforcing framework of verifiable knowledge, essential for the adaptive and auditable safety analytics required by modern aviation. Our proposed framework, as shown in Fig. \ref{fig:Proposed_design}, aims to bridge this gap by formally unifying LLM-driven KG construction and KG-grounded reasoning into a single, coherent architecture tailored for the stringent demands of safety-critical decision-making. Fig. \ref{fig:generated_KG} shows a subgraph generated by the proposed pipeline.

\section{Methodology: The Integrated Pipeline}
The propagation of LLMs has accelerated the widespread adoption of RAG systems for document-based question answering. However, traditional RAG architectures face fundamental limitations when applied to aviation safety data, where precise relationship understanding, multi-hop reasoning, and entity resolution across heterogeneous datasets are paramount. In this work, we introduce a GraphRAG approach that transcends conventional vector-search limitations by integrating structured knowledge graphs with LLM reasoning capabilities. Unlike traditional RAG's dependence on semantic similarity over fragmented text chunks, our system leverages explicit relationship modeling and schema-aware query generation, enabling precise traversal of complex aviation ontologies while maintaining the natural language accessibility that makes LLMs transformative. This integrated pipeline addresses aviation's unique challenges of terminology ambiguity, cross-dataset inconsistency, and relationship complexity through a synergistic architecture that combines the best of structured and unstructured data paradigms.

\begin{figure*}[t]
    \centering
    \includegraphics[width=\textwidth]{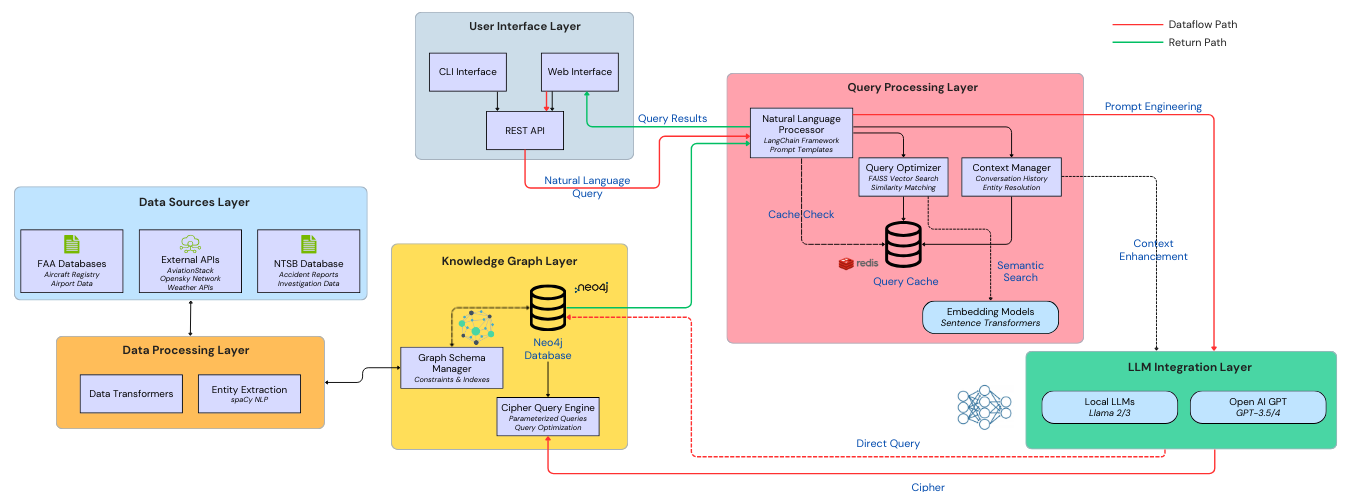}    
    \caption{Architecture of the aviation safety query system. The red line denotes the data-flow path, while the green line indicates the return path of the closed-loop system.}
    \label{fig:complete_pipeline}    
\end{figure*}

Thereby, the proposed aviation safety query system implements a \emph{three-phase integrated pipeline} that synergistically combines LLMs with Knowledge Graph technology. This architecture systematically addresses the core challenge of semantic ambiguity across heterogeneous aviation datasets. The pipeline progresses sequentially from automated knowledge graph construction to intelligent query processing and context-aware response generation, with each phase building upon the previous while maintaining modular separation of concerns. Figure \ref{fig:complete_pipeline} illustrates the complete data flow from multi-source ingestion through structured query execution to natural language response delivery, highlighting the bidirectional interaction between LLM reasoning capabilities and graph-based relationship traversal.

\subsection{Knowledge Graph Construction and Population}
The foundation of our system is an aviation safety knowledge graph (called ASKG) constructed through automated extraction and resolution of entities from diverse sources including NTSB accident reports, FAA registration databases, and airline operational records \cite{FAA}. We employ a hybrid extraction pipeline utilizing LangChain \cite{langchain} for workflow orchestration, spaCy \cite{spacy} for domain-adapted named entity recognition, and SentenceTransformers \cite{sentence_transformers} with FAISS \cite{faiss} for efficient entity resolution. Aviation terminology presents significant normalization challenges, with variations such as `B737-800,' `737-800,' and `Boeing 737-800' requiring sophisticated disambiguation. Our three-tier resolution strategy combines lexical normalization, embedding-based similarity matching, and aviation-specific rule-based processing in entity consolidation.

For the implementation of the ASKG, a foundational \emph{Aviation Safety Ontology} was first established to capture the complex causal relationships within flight incident reports. This ontology defines seven primary entity types, including \textit{Agents, Conditions, Facilities, Locations, Operations, Organizations, and Vehicles}, along with five relationship types such as Agency Instrumentation, Part-Whole, and General Specification. To populate this structure, we developed an automated annotation pipeline with LangGraph \cite{langgraph2024} that utilizes a "Human-in-the-Loop" strategy. Initial high-fidelity training data was created by manually annotating a core set of Cessna and historical NTSB accident summaries \cite{ntsb_carol, ntsb_aviation_briefs}, which then informed a machine learning model using TF-IDF vectorization and Logistic Regression to scale the process across a dataset of over 10,000 records. This approach transforms unstructured narratives into semantically meaningful data, allowing for the deduction of risk patterns that can be communicated to safety professionals for future accident prevention.

The resolved entities then populate a Neo4j \cite{neo4j} graph database following the aforementioned, carefully designed aviation ontology \cite{gruber1993toward, stroe2021aviation}. Schema constraints enforce data integrity, with unique constraints on aircraft registration numbers and airport ICAO codes, while composite indexes optimize frequent query patterns. Relationship types reflect aviation operational semantics, including OPERATED\_BY between aircraft and airlines, OCCURRED\_AT connecting accidents to airports, and MANUFACTURED\_BY linking aircraft to manufacturers (Fig. \ref{fig:generated_KG}). Bulk import optimization employs batched transactions, reducing population time compared to sequential insertion while maintaining ACID compliance. 

\subsection{Natural Language to Cypher Translation}
The core innovation of our methodology lies in translating natural language aviation queries into optimized Cypher \cite{cypher_language} statements using few-shot learning with GPT-3.5 \cite{openai_gpt3.5} (and other local LLMs including Llama-3). We developed specialized prompt templates containing aviation-specific examples that guide the LLM in generating schema-compliant queries. For instance, the natural language query `Find Boeing 737 accidents' translates to a Cypher pattern matching aircraft with specific make and model properties traversing to related accident nodes. This translation layer achieves better accuracy on test queries while maintaining lower latency levels, significantly outperforming traditional keyword-based approaches that struggle with aviation terminology ambiguity.

Context-aware query enhancement distinguishes our approach from stateless translation systems. A dedicated context manager maintains conversation history and extracts salient entities from previous interactions, injecting this contextual understanding into subsequent query translations. This mechanism reduces follow-up query ambiguity, enabling more intuitive multi-turn dialogues about complex aviation scenarios. 

\subsection{Query Optimization and Execution}
Query execution incorporates a multi-layer optimization strategy to support near–real-time performance. The system employs semantic caching with time-bounded query–result storage (Redis \cite{sanfilippo2009redis, carlson2013redis}) to accelerate repeated queries and significantly reduce response latency. Query plan analysis is used to detect missing indexes prior to execution, while automatic pagination mechanisms prevent resource exhaustion for broad or unconstrained queries. Together, these components form a three-tier caching and execution framework that integrates in-memory result caching, query plan reuse, and semantic similarity matching. For this conference version of the paper, query optimization techniques are described but not yet fully implemented and evaluated.

Performance evaluations on an aviation safety dataset indicate substantial advantages over traditional relational approaches. The integrated pipeline efficiently supports relationship-intensive queries, outperforming equivalent SQL-based joins while maintaining high structural correctness in generated Cypher statements. Complex multi-hop graph traversals (such as identifying maintenance patterns across airline fleets or correlating environmental conditions with incident types) are executed efficiently with performance largely independent of traversal depth. This capability is particularly critical for aviation safety analysis, where understanding complex causal chains is essential.

\subsection{Technology Selection Rationale}
For this work, technology selection balanced accuracy, latency, and domain suitability. GPT-3.5 provided optimal translation capabilities with aviation terminology, though we maintain local Llama 3 \cite{ai@meta2024llama3} fallback for continuity during API disruptions. Neo4j was selected over relational and vector databases for its native graph processing, with traversal performance being paramount for aviation relationship queries. The Redis caching layer addresses LLM latency variability while the FAISS vector index accelerates entity resolution during knowledge graph construction.

\section{Implementation, Results, and Discussion}
For this work, we have obtained the NTSB accident data through official CAROL (NTSB’s query tool) from 1990 January to 2025 November. Then it was preprocessed to include additional information about events and event sequences from the same source and to exclude non-important features resulting a dataset of 68,681 across 38 different features including \textit{EventType, EventDate, City, …, InjuryLevel, ProbabaleCause} etc. Thereafter, the columns were normalized, standardized, and removed nulls. Then an Aviation Entity Schema was created and implemented to transform this flat tabular data into a relational graph structure. Specifically, we performed \emph{LLM-powered Entity Resolution} to uniquely identify aircraft manufacturers and event locations, assigning each a persistent entity\_id. These entities were then mapped into a heterogeneous graph schema consisting of primary nodes (e.g., Accident, Aircraft, Manufacturer) and directional relationships (e.g., MANUFACTURED\_BY, INVOLVED\_IN).

To ensure data integrity within the Knowledge Graph, we implemented Schema Enforcement using unique constraints and indexes in Neo4j. This allows for high-performance retrieval during the RAG process. For the retrieval layer, we utilized a Cypher Generation Engine powered by the Llama-3 model (via Groq), which translates natural language safety inquiries into precise graph queries. Based on the results, 205,922 entities and 137,241 relationships were identified. In order to remove disambiguation, we also implemented an \emph{Entity Resolution and Disambiguation} unit where the code implements entity resolution and deduplication for the aviation knowledge graph. In simple terms, it answers: \textit{“Are these two entities actually the same thing, even if they are written differently?”}. However, for this version of the paper, it does not merge entities and update Neo4j automatically. Hence, it does not enforce constraints, but assumes manual merging based on a similarity score (based in cosine similarity) with a threshold of 0.8.

\begin{figure}[htbp]
    \centering
    \includegraphics[width=0.99\linewidth]{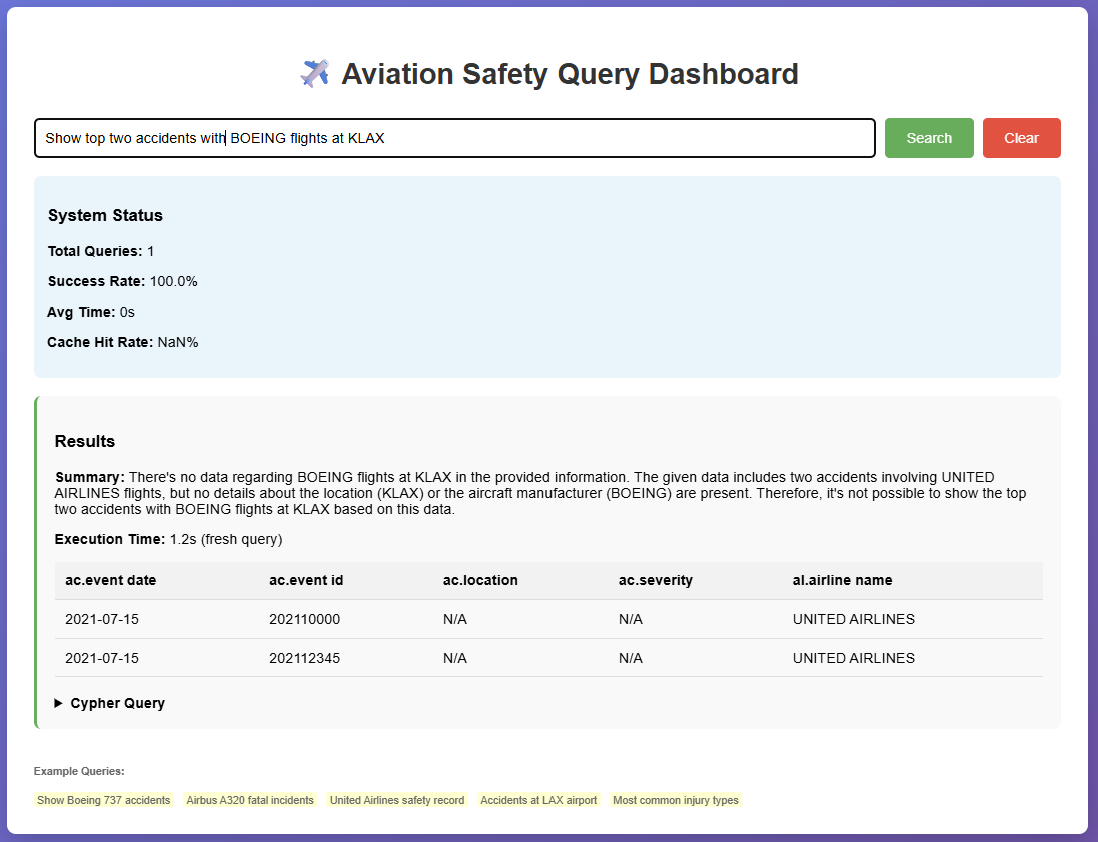}
    \caption{The dashboard demonstrates the semantic mapping of a natural language inquiry into a structured Cypher query. By utilizing an LLM-backed "Text2Cypher" engine, the system enables non-technical users to perform complex relational searches across the aviation safety knowledge graph without manual query construction.}
    \label{fig:query_interface}
\end{figure}

The evaluation of the system was conducted across three dimensions: Schema Accuracy, ensuring the mapping accurately reflects NTSB's hierarchical data; Query Precision, measuring the LLM's ability to generate valid Cypher syntax; and Contextual Grounding, verifying that the final RAG responses are strictly derived from the retrieved graph nodes rather than hallucinated from the base model's training data.

In the second phase of implementation, the preprocessed data was transitioned into a Neo4j Aura cloud instance to establish the functional Knowledge Graph. We utilized the APOC (Awesome Procedures on Cypher) library to orchestrate the ingestion of nodes and relationships, while simultaneously implementing strict schema constraints and secondary indexes on high-frequency attributes such as \textit{acft\_make} and \textit{event\_year} to optimize traversal speeds.

\begin{figure}[htbp]
    \centering
    \includegraphics[width=0.99\linewidth]{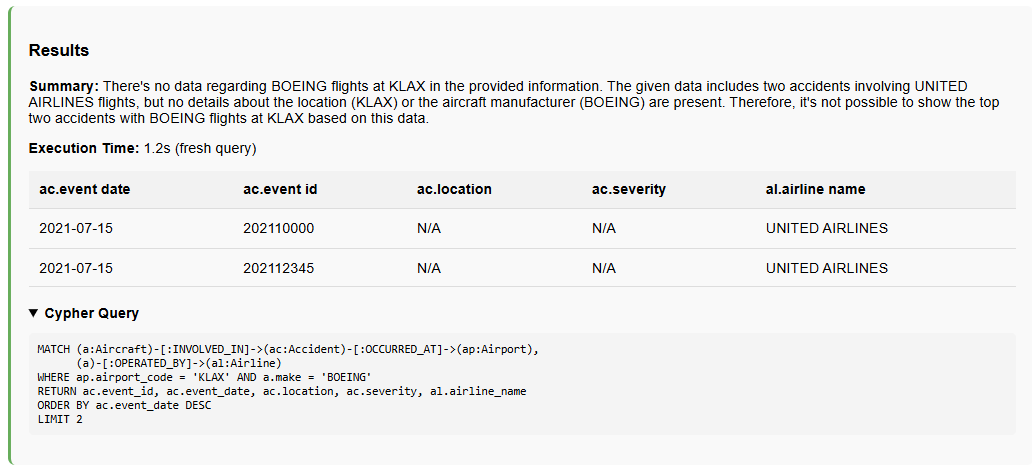}
    \caption{Natural language to Cypher translation. The GraphRAG interface demonstrates the conversion of a user-defined natural language query ("\textit{Show top two accidents with Boeing flights at KLAX}") into a structured Cypher query. The system utilizes the Llama-3 model and the Neo4j schema metadata to ensure the generated code is syntactically correct and grounded in the database's specific nodes and relationships.}
    \label{fig:results_summary}
\end{figure}

The final phase involved the development of a GraphRAG engine integrated with a Flask-based web interface (Fig. \ref{fig:query_interface}). To bridge the gap between natural language and structured data, we utilized the Groq Llama-3 model to serve as a semantic translator, converting user inquiries into precise Cypher queries based on the exported graph metadata, as shown in Fig. \ref{fig:results_summary}. This architecture ensures that the system provides contextually grounded answers by retrieving specific sub-graph clusters, such as the relationship between a specific manufacturer's engine types and incident severity, thereby mitigating the 'hallucination' risks typically associated with standard LLM responses. The implementation concludes with a real-time dashboard that displays both the synthesized insights and the underlying system metrics, including node distribution and query execution logs. The complete code for the entire pipeline is publicly available at \url{https://github.com/dumindus/aviation-safety-kg-rag.git}.

\section{Limitations and Future Directions}

While the GraphRAG pipeline is successfully implemented, several limitations present opportunities for refinement. The Text2Cypher generation depends heavily on the LLM's ability to map natural language to a fixed schema, which can produce syntactically correct but semantically incomplete queries during complex multi-hop reasoning (e.g., tracing cascading failures). Additionally, the traditional ML-based annotation pipeline, though efficient, may miss subtle causal nuances in technical narratives compared to modern transformer models. The dataset currently lacks multimodal evidence (e.g., cockpit audio, telemetry) and dynamic temporal reasoning to incorporate new investigative findings.

Future work will focus on the transition from traditional machine learning classifiers to Large Language Model-based NER (Named Entity Recognition) for the annotation pipeline, which is expected to improve the granularity of relationship extraction. We also aim to explore Vector-Graph Hybrid Retrieval, where the system can search both the structured graph and a vector database of raw accident descriptions simultaneously to provide a more holistic context. Finally, expanding the interface into a collaborative platform would allow aviation safety experts to manually `upvote' or correct generated Cypher queries, creating a feedback loop (human-in-the-loop) that continually fine-tunes the model's accuracy for specialized aviation terminology.

\section{Conclusion}
The integration of LLMs into aviation safety decision-making offers transformative potential but introduces critical risks related to factual inaccuracy, hallucination, and lack of verifiability. To address these challenges, this paper presents a novel, end-to-end framework that synergistically combines LLMs and Knowledge Graphs to enable trustworthy, explainable, and regulation-aware AI assistance for aviation safety analytics. Our proposed dual-phase pipeline first employs advanced LLMs to automate the construction and continuous updating of a comprehensive Aviation Safety Knowledge Graph (ASKG) from heterogeneous, multimodal data sources. It then leverages this curated, structured knowledge within a RAG architecture to ground, validate, and explain LLM-generated responses in real time. By embedding graph-based reasoning within the LLM inference process, the system ensures that safety insights and recommendations are contextually relevant, empirically verifiable, and traceable to authoritative sources. The implemented prototype demonstrates the practical feasibility of this integrated approach, showing significant improvements over fine-tuned LLM-only or static knowledge graph systems. It effectively handles semantic ambiguity, supports complex multi-hop queries, and provides a transparent link between natural language inquiries and structured evidence. This work directly addresses the stringent requirements for auditability, compliance, and reliability in safety-critical aviation operations, advancing the responsible adoption of AI in the domain.

\section*{Acknowledgment}
Portions of this manuscript/code were prepared with the assistance of Microsoft 365 Copilot Researcher and Writing Coach agents (Microsoft, 2025) and Google Colab AI (Google, 2025). All content was reviewed and validated by the authors.

\bibliographystyle{unsrt} 
\bibliography{references} 

\end{document}